\documentclass[twocolumn,aps,prd,showpacs,preprintnumbers,nofootinbib]{revtex4}
\usepackage{amsfonts,amssymb,amsmath}  
\usepackage{graphicx}  
\usepackage{subfigure}  
\usepackage{wrapft}  
\usepackage{color}  
\usepackage{bm}
\newcommand{\f}[2]{\frac{#1}{#2}}  
\newcommand{\mk}[1]{\left( #1 \right)}  
\newcommand{\kk}[1]{\left[ #1 \right]}  
  
\newcommand{\be}{\begin{equation}}  
\newcommand{\ee}{\end{equation}}
\newcommand{\bea}{\begin{eqnarray}}  
\newcommand{\eea}{\end{eqnarray}}
\newcommand{\Mpl}{M_{\rm Pl}}
\newcommand{\ini}{{\rm ini}}

\newcommand{\ie}{\textsl{i.e.~}}

\def\nn{\nonumber}

\def\l{\left}
\def\r{\right}
\def\cR{{\zeta}}
\def\d{{\rm d}}
\def\cG{{\cal G}}
\def\cR{{\zeta }}
\def\cB{{\cal B}}
\def\fnl{f_{_{\rm NL}}}

\def\vk{{\bm k}}
\def\vka{{\bm k}_{1}}
\def\vkb{{\bm k}_{2}}
\def\vkc{{\bm k}_{3}}
\def\ka{{\bm k}_{1}}
\def\kb{{\bm k}_{2}}
\def\kc{{\bm k}_{3}}
\def\ei{\eta_{\rm i}}
\def\ef{\eta_{\rm f}}

\def\fnl{f_{_{\rm NL}}}

\begin{document}
\preprint{RESCEU-47/12}  

\title{  
Ultra Slow-Roll Inflation and the non-Gaussianity Consistency Relation
}

\author{J\'er\^ome Martin$^{~1,2}$}
\email[E-mail:]{jmartin@iap.fr}
\author{Hayato~Motohashi$^{~2,3}$}  
\email[E-mail:]{motohashi@resceu.s.u-tokyo.ac.jp}  
\author{Teruaki Suyama$^{~2}$}  
\email[E-mail:]{suyama@resceu.s.u-tokyo.ac.jp}  
\address{  
$^{1}$ Institut d'Astrophysique de Paris, UMR 7095-CNRS,
Universit\'e Pierre et Marie Curie, 98 bis boulevard Arago, 75014
Paris, France \\
$^{2}$ Research Center for the Early Universe (RESCEU),  
Graduate School of Science, The University of Tokyo, Tokyo 113-0033, Japan \\
$^{3}$ Department of Physics, Graduate School of Science,  
The University of Tokyo, Tokyo 113-0033, Japan 
}  
  
\begin{abstract}
Ultra slow-roll inflation has recently been used to challenge 
the non-Gaussianity consistency relation. We show that this 
inflationary scenario belongs to a one parameter class of models and we study its 
properties and observational predictions. We demonstrate that the 
power spectrum remains scale-invariant and that the bi-spectrum 
is of the local type with $\fnl=5(3-n_{_{\rm S}})/4$ which, indeed, 
represents a modification of the consistency relation. However, we 
also show that the system is unstable and suffers from many physical 
problems among which is the difficulty to correctly 
WMAP normalize the model.
We conclude that ultra slow-roll inflation remains 
a very peculiar case, the physical relevance of which is probably 
not sufficient to call into question the validity of the consistency 
relation.  
\end{abstract}
\pacs{98.80.Cq}  
\maketitle  

\section{Introduction}
\label{sec:intro}

The theory of inflation convincingly describes the 
physical conditions that prevailed in the very early 
Universe~\cite{Martin:2006rs,Lorenz:2007ze,Lorenz:2008je,
Martin:2010kz,Martin:2010hh}. However, there are many models of inflation 
and it is not yet clear which scenario is actually realized
in Nature. For this reason, the recent developments 
in the calculations of higher correlation functions~\cite{Maldacena:2002vr} are 
important since they might allow us to constrain and maybe 
rule out many models of inflation. For instance, the simplest
scenarios (\ie a slowly rolling single field with a canonical 
kinetic term) are known to predict a negligible level of 
non-Gaussianity, of the order of the slow-roll parameters~\cite{Gangui:1993tt,Gangui:1994yr,Wang:1999vf,Gangui:1999vg,Gangui:2000gf,Gangui:2002qc}. 
Therefore, if any non-Gaussianity is detected in the future (for instance with the Planck satellite), these models would be excluded.

\par

Recently, however, it was argued in Ref.~\cite{Namjoo:2012aa} 
that this is not necessarily true for the simple scenarios
mentioned above. An explicit counter-example was 
investigated in Ref.~\cite{Namjoo:2012aa} and it was shown that, in this 
particular case, the value of the $\fnl$ parameter can 
be a few instead of being negligible. Since this result
challenges a well-known and important theorem, it is important to study 
in more detail the model that has been utilized to obtain this conclusion. In 
particular, one would like to know whether this just represents a 
very peculiar case or whether this can correspond to a 
generic class of meaningful models. 

\par

In fact, the inflationary scenario used in Ref.~\cite{Namjoo:2012aa} has 
been known for a long time and is named "ultra slow-roll" 
inflation. It was studied for the first time in Ref.~\cite{Kinney:2005vj} 
(Similar situations were also investigated in Ref.~\cite{Inoue:2001zt}).
In the present article, we show that it belongs to a broader 
class of models that we explicitly identify. 
The goal of the paper is then to study this new family of scenarios, their 
properties and the 
corresponding observational predictions (power spectrum and 
bi-spectrum).

\par

The paper is organized as follows. In the next section, 
Sec.~\ref{sec:ultra}, we introduce ultra slow-roll inflation 
and show how it can be generalized. Then, we study the stability 
of the system and investigate whether one can easily produce 
$60$ e-folds in the ultra slow-roll regime. Then, in Sec.~\ref{sec:usrps}, 
we calculate the power spectrum of curvature fluctuations and show 
that it can be scale invariant even if the slow-roll parameters 
are not all small. In Sec.~\ref{sec:usrNG}, we estimate the non-Gaussianities 
and show that the Maldacena's consistency relation is indeed violated. As 
a consequence the $\fnl$ parameter is of order one in this class of 
models. Finally, in Sec.~\ref{sec:conclusion}, we discuss in more 
details the difficulties of ultra slow-roll inflation and present 
our conclusions.

\section{Ultra Slow-roll Inflation}
\label{sec:ultra}

\begin{figure*}
\includegraphics[width=0.45\textwidth,height=.45\textwidth]{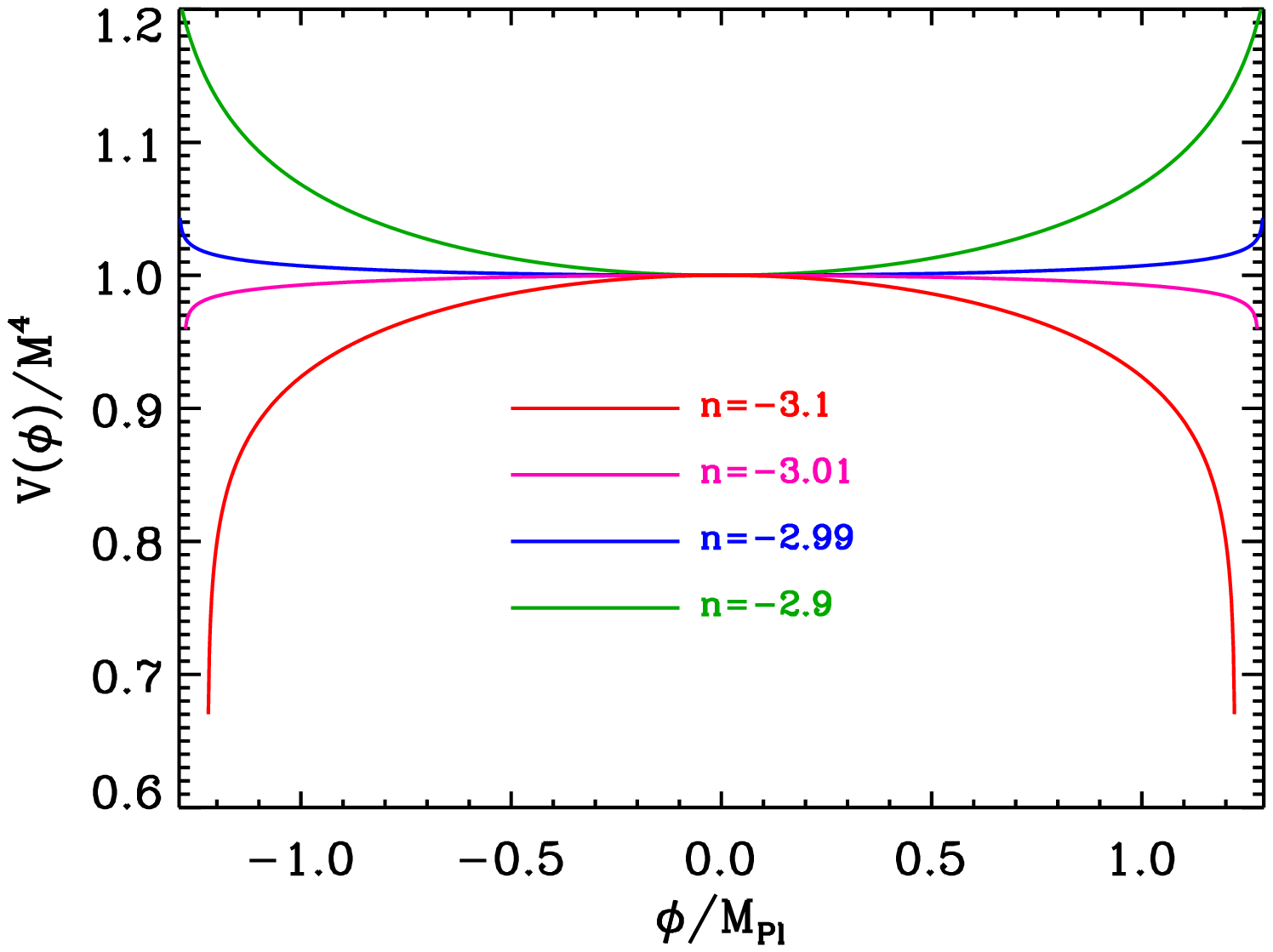}
\includegraphics[width=0.45\textwidth,height=.45\textwidth]{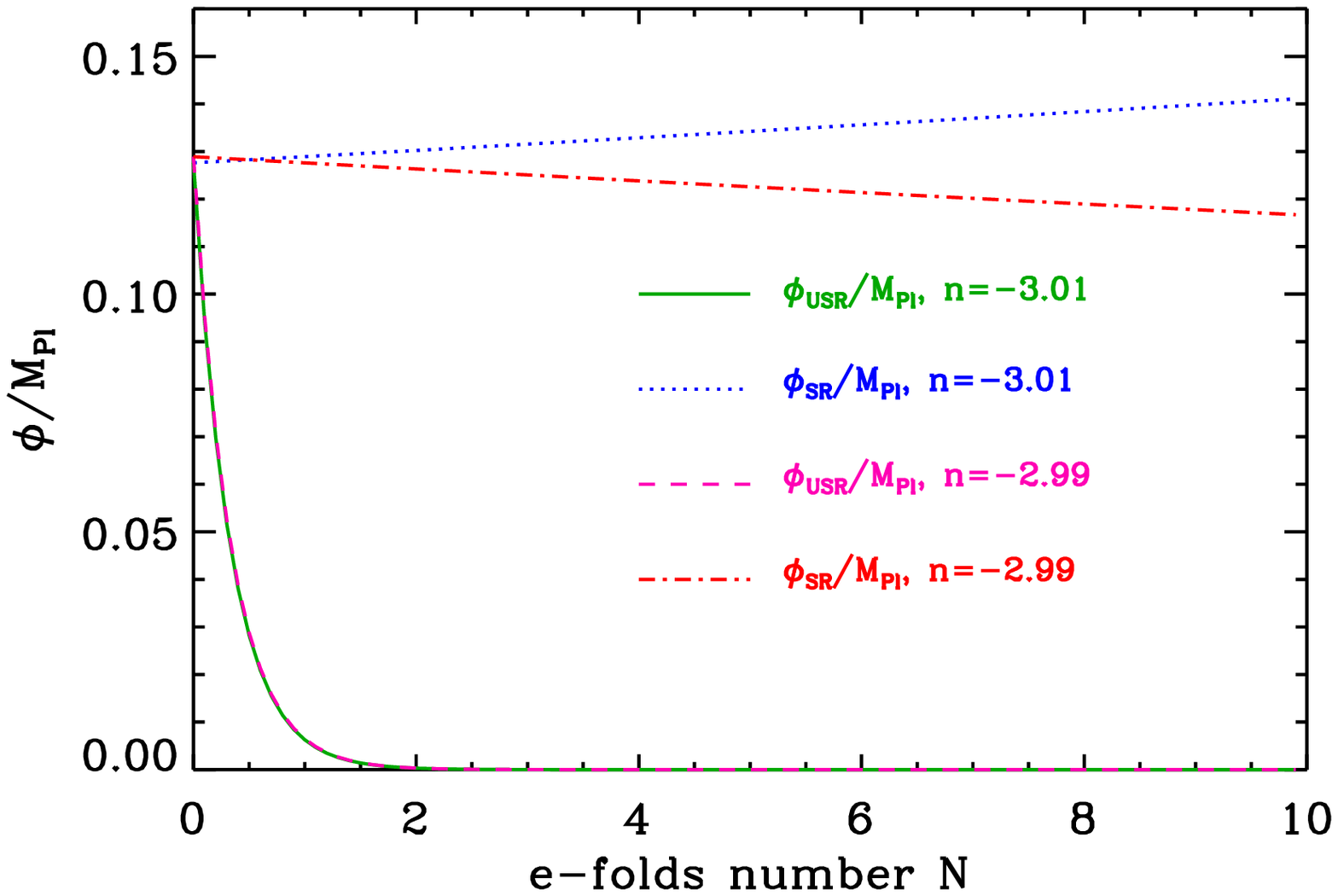}
\caption{Left panel: new family of ultra slow-roll potentials for different values of the parameter $n$.
Right panel: classical ultra slow-roll and slow-roll trajectories for $n=-3.01$ (solid green line and dotted blue 
line) and $n=-2.99$ (dashed pink line and dotted dashed red line). The initial condition for 
the scalar field is chosen to be $\phi_{\rm ini}=0.1\, \phi_{\rm lim}$.}
\label{fig:potential}
\end{figure*}

Let us consider an inflationary model with a single scalar 
field $\phi$ (with a standard 
kinetic term). The equations of 
motion for $\phi $ and for the Friedmann-Lema\^itre-Robertson-Walker scale factor 
$a(t)$ ($t$ denotes the cosmic time and, in the following, 
a dot means a derivative with respect to $t$) are the Friedmann and the 
Klein-Gordon equations, namely
\begin{eqnarray}
3 \Mpl^2 H^2 &=& \f{\dot \phi^2}{2}+V(\phi), \\
\ddot \phi + 3H\dot \phi + V_\phi&=& 0, 
\end{eqnarray}
where $H=\dot{a}/a$ is the Hubble parameter and $\Mpl$ the reduced Planck mass. The 
background evolution can also be characterized in terms of the 
slow-roll parameters (or horizon-flow parameters) defined by
\begin{equation}
\epsilon_{i+1}=\frac{{\rm d}\ln \epsilon_i}{{\rm d}N},
\end{equation}
where $N$ denotes the number of e-folds, $N\equiv \ln (a/a_{\rm ini})$ ($a_{\rm ini}$ being the scale factor at the beginning of inflation). The hierarchy starts with $\epsilon_0\propto 1/H$ which implies that the first slow-roll parameter can be expressed as 
\be \epsilon_1\equiv -\f{\dot H}{H^2} 
=\frac{\dot{\phi}^2}{2\Mpl^2H^2}.\ee
The second slow-roll parameter can be used to express the acceleration of the field, namely
\begin{equation}
\epsilon_2=2\epsilon_1+2\frac{\ddot \phi}{H\dot \phi}.
\end{equation}
Inflation requires $\epsilon_1<1$ and the slow-roll approximation 
is valid if all the horizon-flow parameters are small, $\epsilon_i\ll 1$ 
during inflation.

\par

As discussed in the introduction, if the potential is exactly flat, then 
the Klein-Gordon equation implies that $\ddot{\phi}/(H\dot \phi)=-3$ and this corresponds to the situation discussed in Ref.~\cite{Kinney:2005vj} and named "ultra slow-roll inflation". In this case, despite the flatness of the potential, the slow-roll parameters are not all small: usually $\epsilon_1 \ll 1$ but obviously $\epsilon_2 ={\cal O}(1)$. As a consequence, one could expect the power spectrum to deviate from scale-invariance but, as shown in Ref.~\cite{Kinney:2005vj}, and as discussed in more detail in the next section, this is in fact not the case. This makes this model a priori interesting since this shows 
that scale invariance can be obtained even if the slow-roll approximation is violated. This also raises the question of whether this is peculiar to the property $V_\phi=0$ or whether this can also be obtained in a broader context. In order to investigate this issue let us consider the more general condition
\begin{equation}
\label{eq:ultracond}
\ddot\phi=nH\dot\phi, 
\end{equation}
where $n$ is now an arbitrary number, not necessarily equal to $-3$. Let us also assume that the first slow-roll parameter is still very small, $\epsilon_1 \ll 1$. Obviously, the case $n\simeq 0$ corresponds to slow-roll and $n= -3$ to ultra slow-roll. The corresponding equations of motion are given by
\bea
\label{eq:friedmanusr}
3 \Mpl^2 H^2 &\simeq & V(\phi), \\
\label{eq:kgusr}
(n+3)H\dot\phi+V_\phi&=& 0.
\eea
From these equations, it is easy to check that $\dot{\phi}\propto a^n$, which implies
that 
\begin{equation}
\epsilon_1\propto a^{2n}, \quad \epsilon_2\simeq 2n, \quad \epsilon_3 =0.
\label{eq:ultrasrparam}
\end{equation}
In particular, for $n=-3$, one recovers the well known scaling 
$\epsilon_1\propto a^{-6}$, see Ref.~\cite{Kinney:2005vj}. 

\par

\begin{figure*}
\includegraphics[width=0.45\textwidth,height=.45\textwidth]{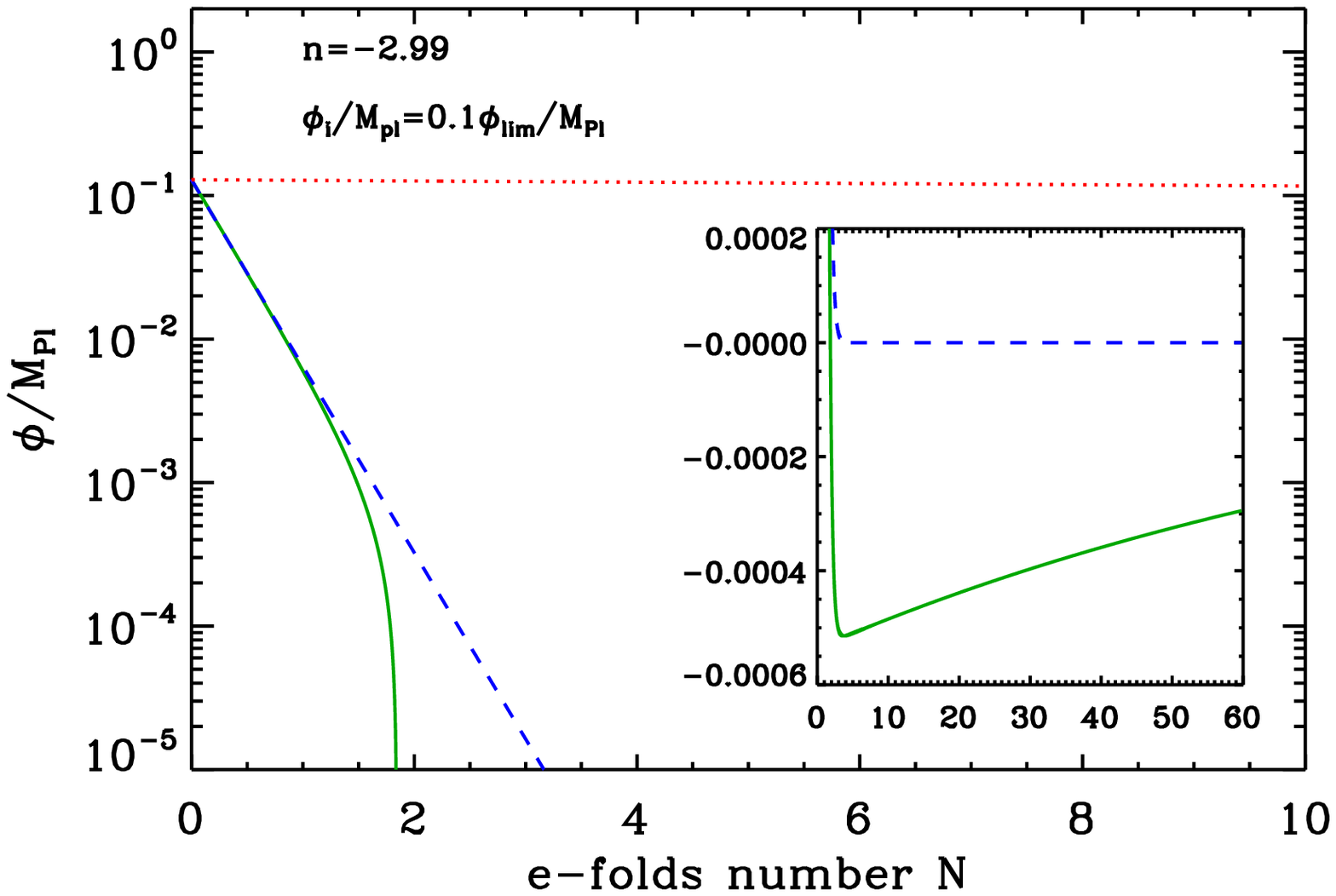}
\includegraphics[width=0.45\textwidth,height=.45\textwidth]{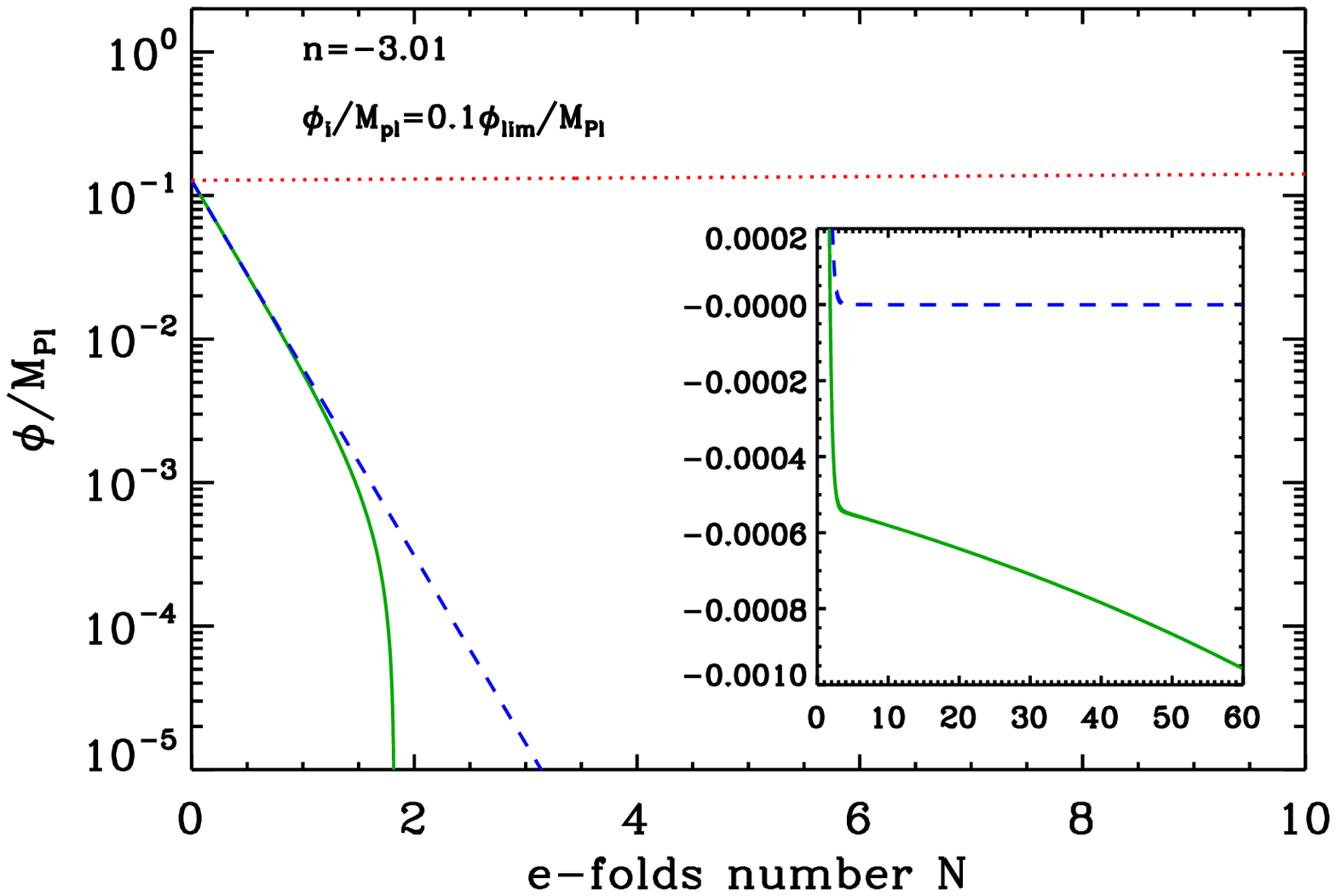}
\caption{Left panel: exact (numerical) evolution of the field (solid green line) 
compared to the ultra slow-roll solution (dashed blue line) and to the slow-roll 
solution (dotted red line). The parameter $n$ is taken to be $n=-2.99$ and the 
initial condition is the same as in Fig.~\ref{fig:potential} (right panel), 
namely $\phi_{\rm ini}=0.1\, \phi_{\rm lim}$. The inset shows the global evolution 
of the system on a larger time scale. Right panel: same as left panel but for $n=-3.01$.}
\label{fig:field}
\end{figure*}

From the two equations of motion~(\ref{eq:friedmanusr}) and~(\ref{eq:kgusr}), it 
is also straightforward to integrate the classical trajectory. One obtains
\begin{equation}
\label{eq:Nphi}
N(\phi)=-\frac{n+3}{3\Mpl^2}\int _{\phi_{\rm ini}}^{\phi}
\frac{V}{V_\phi}{\rm d}\phi,
\end{equation}
where $\phi_{\rm ini}$ denotes the initial value of the inflaton. However, it is not obvious that this solution will satisfy the condition~(\ref{eq:ultracond}). Requiring that this is the case, we find that the potential must obey the following differential equation
\begin{equation}
 \f{V_{\phi\phi}}{V}-\f{3}{2(n+3)}\mk{\f{V_\phi}{V}}^2+\f{n(n+3)}{3\Mpl^2}=0. 
\end{equation}
Interestingly enough, this differential equation can be integrated and leads to the following potential
\begin{equation} 
\label{eq:usrpot}
V(\phi)=M^4\kk{\cos\mk{\sqrt{\f{n(2n+3)}{6}}\f{\phi-\phi_0}{\Mpl}}}^{\f{2(n+3)}{2n+3}}, 
\end{equation}
where $M$ is an arbitrary mass scale to be fixed by the Wilkinson Microwave Anisotropy Probe (WMAP) normalization and $\phi_0$ an arbitrary constant that, without loss of generality, we can take to be $\phi_0=0$. The potentials in Eq.~(\ref{eq:usrpot}) represent a new family of model depending on one parameter, $n$. These potentials are represented in Fig.~\ref{fig:potential}. If $n<-3$, then they are defined only in the range $-\phi_{\rm lim}<\phi<\phi_{\rm lim}$ with $\phi_{\rm lim}/\Mpl\equiv \pi/2 \sqrt{6/[n(3+2n)]}$. It is clear that if $n\simeq -3$, the potential is extremely flat, justifying the name "ultra slow-roll".

\par

\begin{figure*}
\includegraphics[width=0.45\textwidth,height=.45\textwidth]{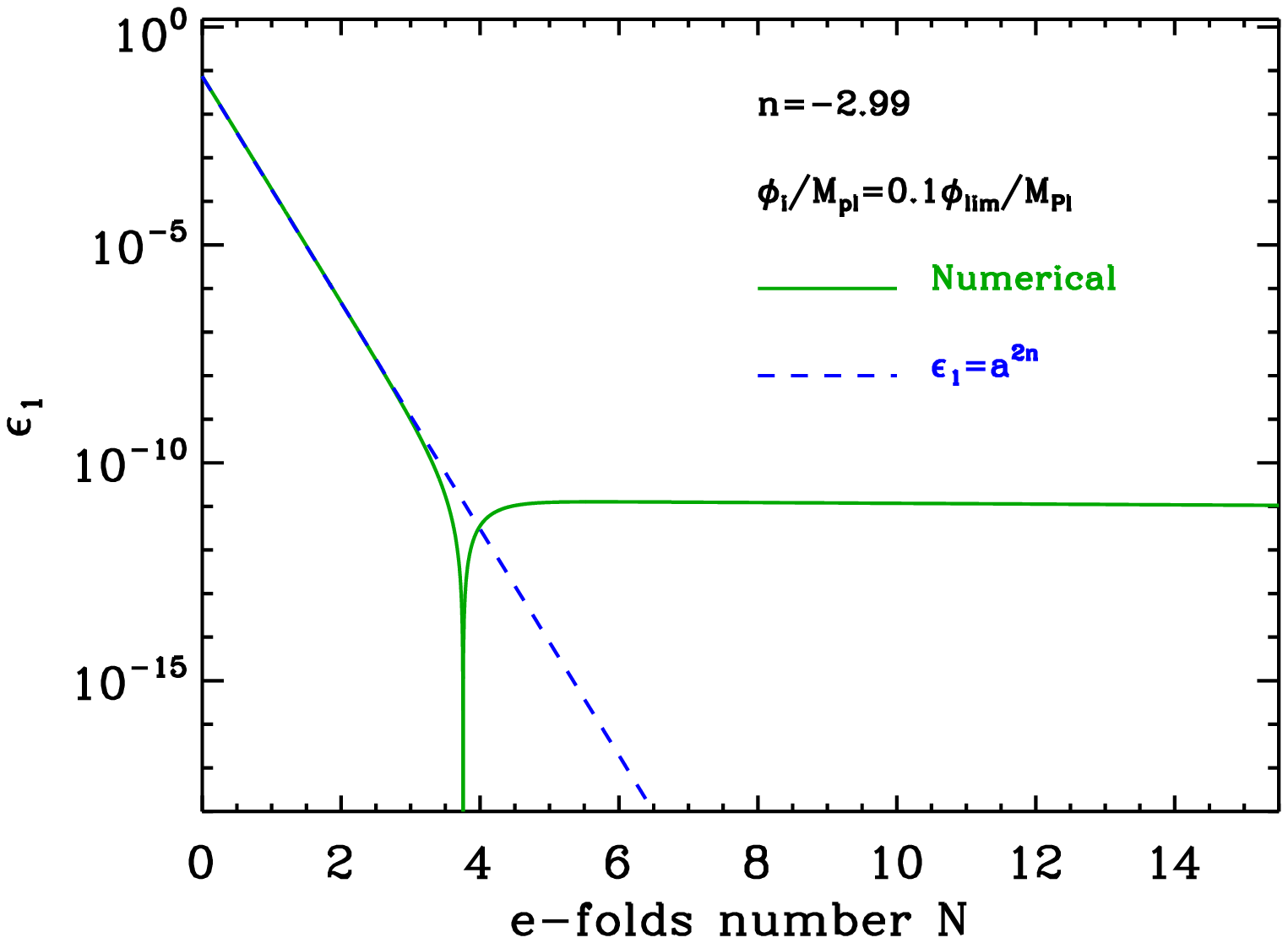}
\includegraphics[width=0.45\textwidth,height=.45\textwidth]{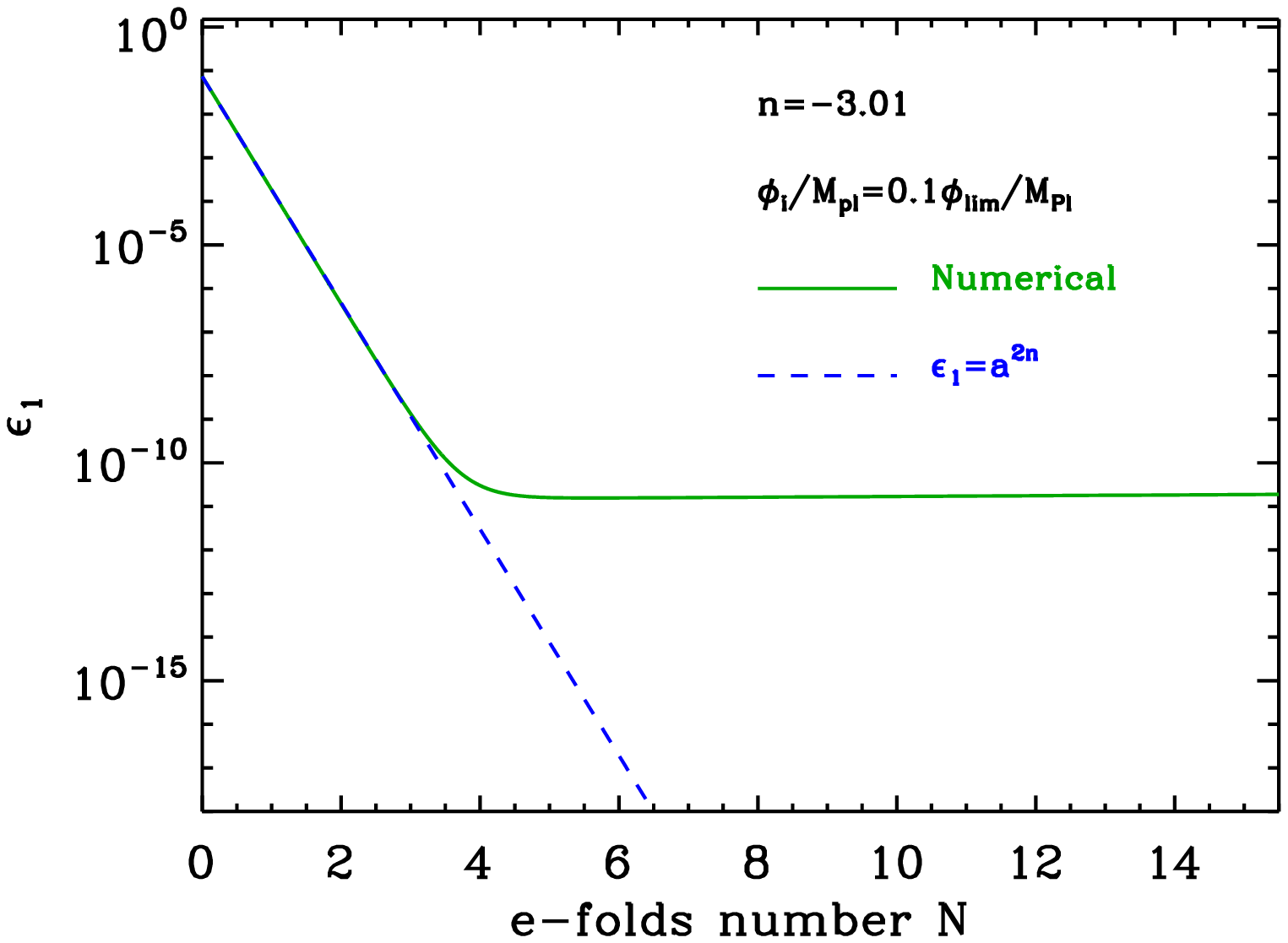}
\includegraphics[width=0.45\textwidth,height=.45\textwidth]{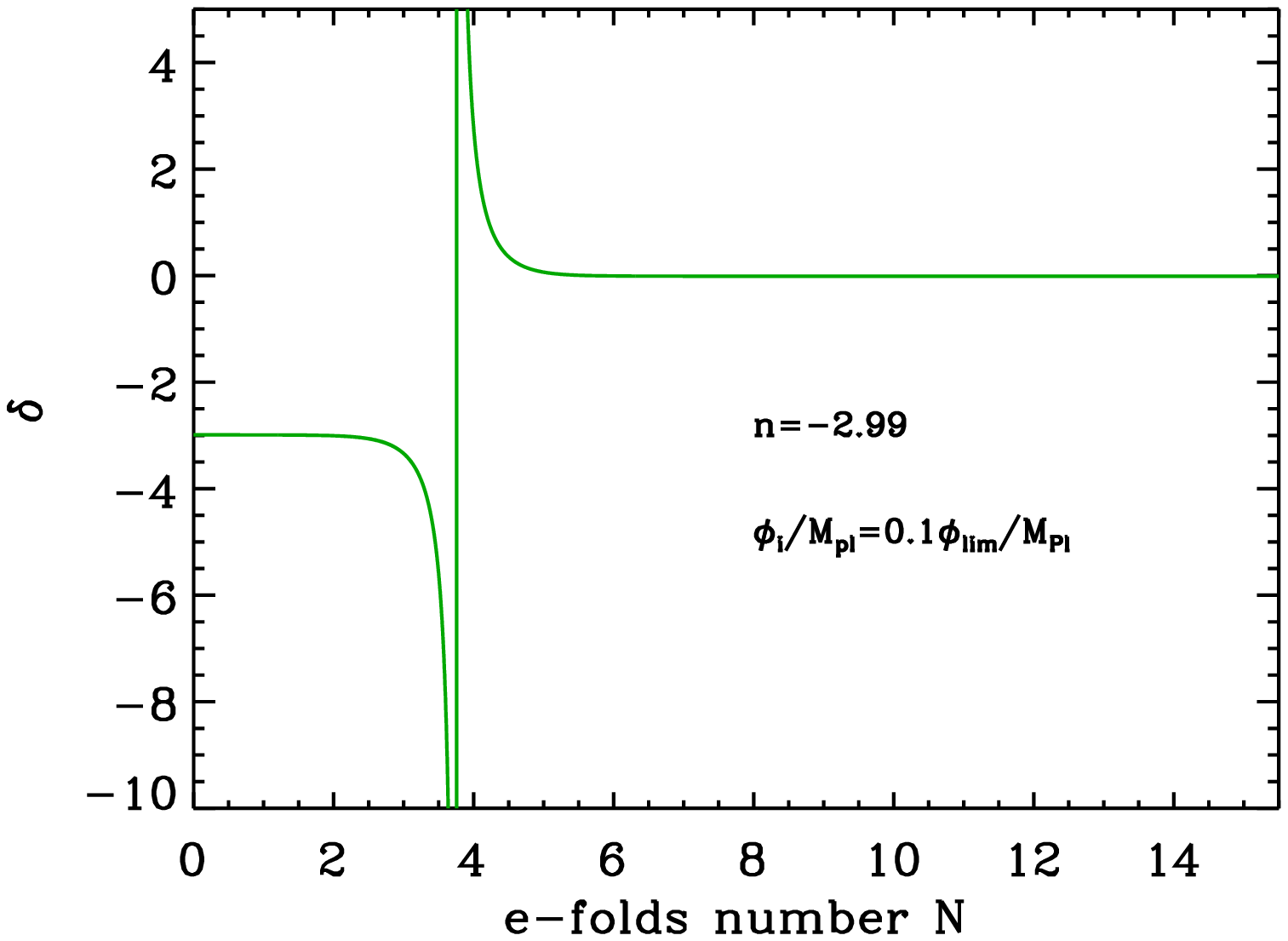}
\includegraphics[width=0.45\textwidth,height=.45\textwidth]{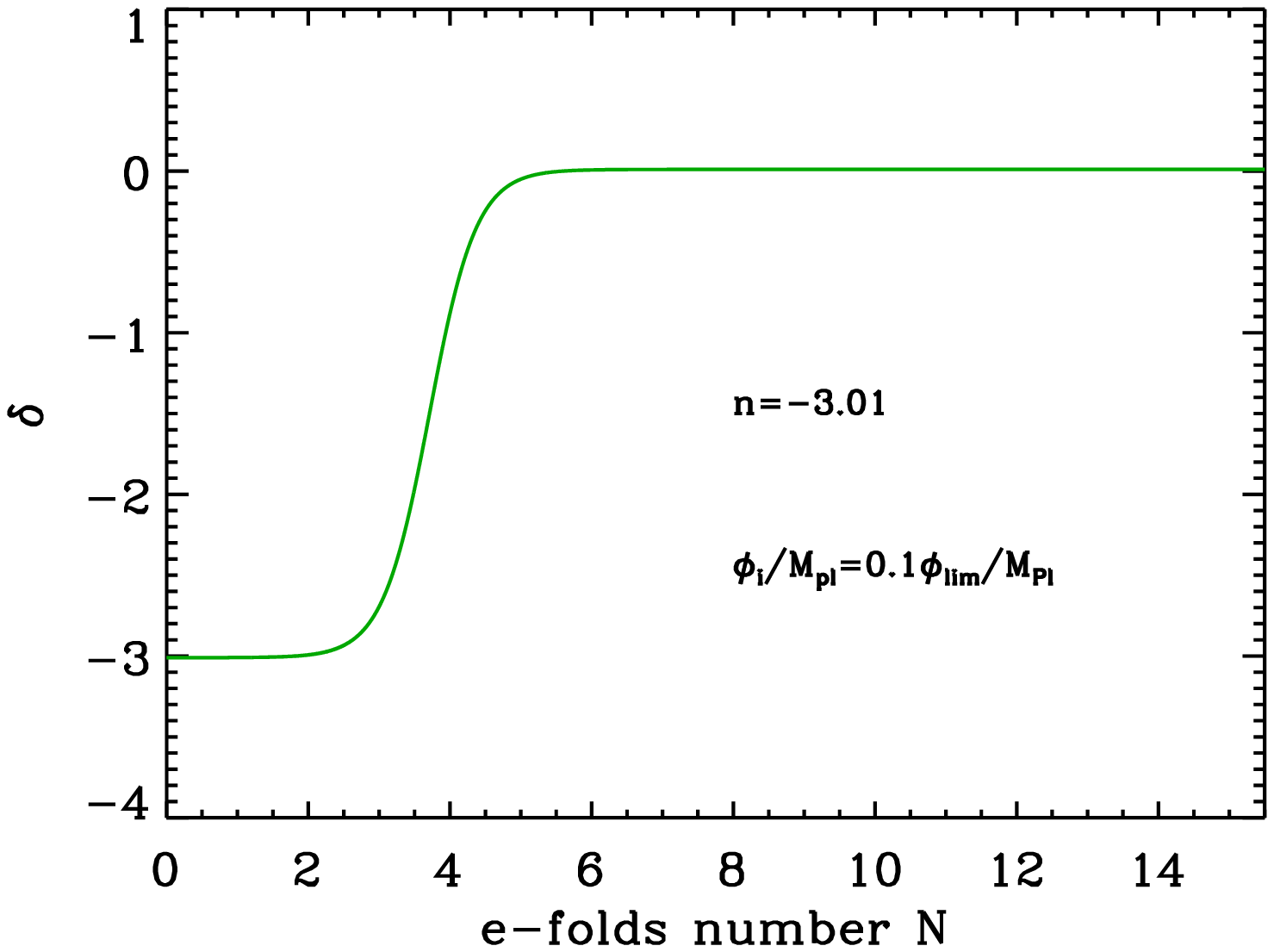}
\caption{Top left panel: numerical (exact) evolution of the first horizon-flow parameter $\epsilon_1$ 
(solid green line) compared to its ultra-slow-roll behavior (dashed blue line) for $n=-2.99$
and $\phi_{\rm ini}=0.1\, \phi_{\rm lim}$. Top right panel: same as top left panel but 
for the choice $n=-3.01$. Bottom left panel: Evolution of the quantity $\delta\equiv \ddot{\phi}/(H\dot{\phi})$ 
for $n=-2.99$ and $\phi_{\rm ini}=0.1\, \phi_{\rm lim}$. Bottom right panel: same as bottom 
left panel but with $n=-3.01$.}
\label{fig:epsilon1}
\end{figure*}

Using Eq.~(\ref{eq:Nphi}), one can compute 
the classical trajectory exactly. Inserting Eq.~(\ref{eq:usrpot}) into Eq.~(\ref{eq:Nphi}) leads to 
the following result
\begin{align}
\phi_{_{\rm USR}}&(N) = \Mpl\sqrt{\f{6}{n(2n+3)}}
\nonumber \\ &\times 
{\rm arcsin}
\left[{\rm e}^{nN}\sin \left(\sqrt{\frac{n(2n+3)}{6}}
\frac{\phi_{\rm ini}}{\Mpl}\right)\right].
\end{align}
One can check that $\phi=\phi_{\rm ini}$ implies $N=0$. Let us notice that this expression 
is not well-defined for the slow-roll case $n=0$. In this situation, one should use the following expression 
\begin{align}
\phi_{_{\rm SR}}&(N) = \Mpl\sqrt{\f{6}{n(2n+3)}}
\nonumber \\ &\times 
{\rm arcsin}
\left[{\rm e}^{n(n+3)N/3}\sin \left(\sqrt{\frac{n(2n+3)}{6}}
\frac{\phi_{\rm ini}}{\Mpl}\right)\right].
\end{align}
The ultra slow-roll and slow-roll trajectories are represented in Fig.~\ref{fig:potential}. 
The interpretation of these results can be easily understood. The slow-roll solutions 
just follow the curvature of the potential. Therefore, if $n\lesssim -3$ then the vacuum 
expectation value of the field increases (the field escapes at infinity and will meet the singularity 
at $\phi=\phi_{\rm lim}$) and inflation proceeds from the left to 
the right while, if $n\gtrsim -3$, the field value decreases toward the minimum of the 
potential and inflation proceeds from the right to the left. The ultra slow-roll solutions 
behave in a different manner. Firstly, they are very similar whatever the sign of $n+3$ provided 
$n\simeq -3$ and, secondly, the field always asymptotically approaches the minimum of the 
potential (\ie $\phi=0$). In the case $n\lesssim -3$, this means that the field actually 
climbs up the potential. This is of course due to the fact that, initially, it possesses a 
non-vanishing and non-negligible velocity.

\par

In order to investigate the stability of the system, we have 
also numerically integrated the exact equations of motion. Recall that 
the ultra slow-roll solution has been obtained from the exact 
equations of motion by neglecting the kinetic term in the right hand 
side of the Friedmann equation. This term, although very small, represents 
a perturbation for the ultra slow-roll solution. It is therefore interesting 
to study whether the system can stay in ultra slow-roll during a large 
number of e-folds. The result is presented in Fig.~\ref{fig:field}. After 
a few e-folds the exact solution (solid green line) leaves the ultra slow-roll 
solution (dashed blue line). On a larger time scale (see the insets in 
Fig.~\ref{fig:field}), we see that for $n\gtrsim -3$ (left panel), the field passes 
through the minimum, becomes negative and starts to climb up the potential 
in the region $\phi<0$. Then it reaches a maximum, turns back and decreases 
toward the minimum. Obviously, this evolution is very different from the 
ultra slow-roll one. For $n\lesssim -3$ (see the right panel), the field 
also becomes negative but, since the curvature of the potential is now 
negative, it simply escapes to infinity in the region $\phi<0$. 

\par

It is also interesting to study the behavior of the first slow-roll parameter
and of $\delta \equiv \ddot{\phi}/(H\dot{\phi})$. They are represented in 
Fig.~\ref{fig:epsilon1}. The conclusions obtained before are confirmed. We see 
that $\epsilon_1$ (top panels) scales as $a^{2n}$ only for a few e-folds and then leaves the 
ultra slow-roll regime. For $n=-2.99$ (top left panel), we also notice that $\epsilon_1$ vanishes 
and, of course, this corresponds to the point where, in the region $\phi<0$, the 
field reaches a maximum and turns back. This is confirmed by the fact that, in 
the case $n=-3.01$ (top right panel), the above mentioned behavior never happens. Then, 
after this transitory regime, in both cases, $\epsilon_1$ becomes constant with a 
very small value. The behavior of $\delta $ (bottom panels) can be 
understood in a similar fashion. Initially $\delta \simeq -3$ since the field starts 
from the ultra slow-roll regime. After a few e-folds, this solution is left and, eventually, 
$\delta $ reaches a regime where it remains constant with a very small value. In 
the case $n=-2.99$ (bottom left panel), 
$\delta $ diverges when $\dot{\phi}=0$ while its evolution remains smooth if $n=-3.01$. It 
is clear that having $\epsilon_1$ and $\delta $ (or, equivalently $\epsilon_2$) small and 
constant corresponds to 
nothing but the slow-roll regime. The conclusion of our numerical investigation is 
therefore that the ultra slow-roll regime is unstable and is left after a few e-folds. Then, the 
system simply converges toward the slow-roll solution.

\par

\begin{figure}
\includegraphics[width=0.45\textwidth,height=.45\textwidth]{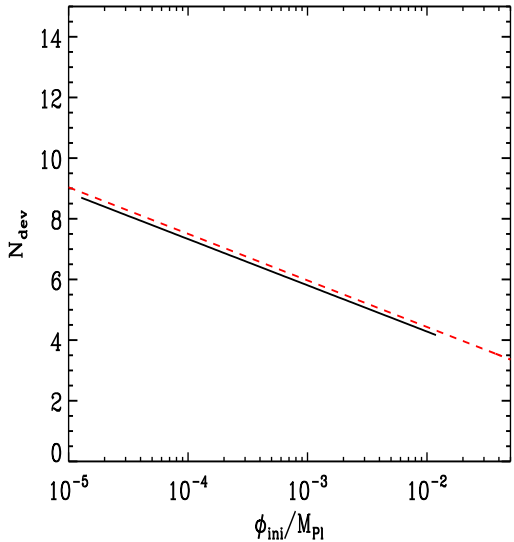}
\caption{Number of e-folds at which the ultra slow-roll solution is left as 
a function of the initial value of the field. The exact numerical result (solid 
black line) is in excellent agreement with the analytical estimate of Eq.~(\ref{eq:Ndev}) 
(dashed red curve).}
\label{fig:deviation}
\end{figure}

It is also interesting to understand when the ultra slow-roll regime 
is left and what are the quantities which control the instability.
We now analyse this question in more detail. For this purpose let 
us define the following quantity
\begin{equation} 
f\equiv\f{\ddot\phi}{nH\dot\phi}=\frac{\delta}{n}, 
\end{equation}
which is one during ultra slow-roll inflation. Using the equations
of motion, it is easy to show that it obeys the following 
first order non linear differential equation
\begin{equation}
\f{{\rm d}f}{{\rm d}N}=-\f{V_{\phi \phi}}{nH^2}+\f{3\epsilon_1}{n}+f(\epsilon_1-nf-3) 
\end{equation}
This equation cannot be solved exactly but we can study the behavior 
of small perturbations. For this reason, we now define $\Delta $ by 
mean of the following formula $f\equiv 1+\Delta $. This quantity obeys the equation
\begin{equation} 
\f{{\rm d}\Delta}{{\rm d}N}
=\f{3+n}{n}\epsilon_1-\Delta\kk{3-\epsilon_1+n(\Delta+2)},
\end{equation}
which, in the regime where $\Delta \ll 1$, can be approximated by
\begin{equation}
\f{{\rm d}\Delta}{{\rm d}N}\simeq \f{3+n}{n}\epsilon_1-\Delta(3+2n) 
\end{equation}
Taking into account the behavior of the first slow-roll parameter during the 
ultra slow-roll regime, namely $\epsilon_1=\epsilon_1\vert_{\rm ini}a^{2n}$, it 
is straightforward to obtain the following solution
\be 
\Delta(N)= \f{n+3}{n(4n+3)}\epsilon_\ini \left[
{{\rm e}^{2nN}-{\rm e}^{-(2n+3)N}}\right].
\ee
For $\vert n+3\vert \ll1$, one can approximate this solution by
\be 
\Delta(N)\simeq-\f{n+3}{27}\epsilon_1\vert_\ini {\rm e}^{3N}.
\ee
This allows us to estimate at which e-folds, $N_{\rm dev}$, 
the actual solution deviates from the ultra slow-roll one.
Straightforward manipulations lead to
\begin{equation}
\label{eq:Ndev}
N_{\rm dev}\simeq \frac23\ln \left(\frac{1}{\vert n\vert}
\sqrt{\frac{54 \Delta _{\rm cri}}{\vert n+3\vert}}
\frac{\Mpl}{\phi_{\rm ini}}\right),
\end{equation}
where $\phi_{\rm ini}$ is the initial value of the field 
and $\Delta_{\rm cri}$ an arbitrary value at which we 
estimate that one has left the ultra slow-roll solution. 
In the following, we estimate that this is the case if the 
actual solution differs for more than $10\%$ from the 
ultra slow-roll one, that is to say $\Delta _{\rm cri}\simeq 0.1$. We 
have computed this quantity numerically and have compared 
it with Eq.~(\ref{eq:Ndev}) in Fig.~\ref{fig:deviation}. Clearly 
the agreement is excellent. The main information brought 
by Eq.~(\ref{eq:Ndev}) is that the dependence in $\phi_{\rm ini}$ 
is logarithmic. The ultra slow-roll solution is interesting if the system 
can follow the corresponding trajectory for at least $60$ e-folds. 
Using Eq.~(\ref{eq:Ndev}), one can estimate what it means for the 
initial conditions. Straightforward manipulations lead to the 
constraint
\begin{equation}
\label{eq:ftic}
\frac{\phi_{\rm i}}{\Mpl}\lesssim \frac{1}{\vert n\vert}
\sqrt{\frac{54 \Delta _{\rm cri}}{\vert n+3\vert}}{\rm e}^{-90}.
\end{equation}
In other words, in order to have $60$ e-folds of ultra slow-roll 
inflation, one must fine-tune dramatically the initial value 
such that it is extremely close to the top of the potential. This 
is of course due to the logarithmic dependence in Eq.~(\ref{eq:Ndev}) which 
is in fact a consequence of the instability of the system.

\section{Ultra Slow-Roll Power Spectrum}
\label{sec:usrps}

The fact that one of the slow-roll parameters is not small 
immediately raises the question as to whether the model can lead to an 
almost scale invariant power spectrum. To address this question, 
it is convenient 
to work in terms of the so-called Mukhanov-Sasaki variable $v_{\bm k}$, 
which is related to the curvature perturbation by
$\zeta_{\bm{k}}=v_{\bm k}/(\sqrt{2}\Mpl a\sqrt{\epsilon_1})$. The
spectrum of $\zeta _{\bm k}$ can be expressed as
\begin{equation}
{\cal P}_{\zeta}(k)\equiv \frac{k^3}{2\pi ^2}
\left\vert \zeta_{\bm k}\right\vert ^2
=\frac{2k^3}{8 \pi^2\Mpl^2}\left\vert 
\frac{v_{\bm k}}{a\sqrt{\epsilon _1}}\right\vert ^2\, .
\label{Pzeta}
\end{equation}
The variable $v_{\bm k}$ obeys the
equation of a parametric oscillator, the time-dependent frequency 
being determined by the dynamics of the background \cite{Mukhanov:1990me}
\begin{equation}
v_{\bm k}''+\left(k^{2}
-\frac{z''}{z}\right)v_{\bm k}=0 ,
\label{eq:eomv}
\end{equation}
where the prime denotes a derivative with respect to conformal time and
where $z$ is given by $a\sqrt{\epsilon_1}$. The quantity $k$ represents the comoving wavenumber of a Fourier mode. The "effective potential" $z''/z$ can be 
expressed as 
\begin{eqnarray}
\frac{z''}{z} &=&a^2 H^2\biggl(2-\epsilon_1 + \frac{3}{2}\epsilon_2 + \frac{1}{4}\epsilon_2^2 - \frac{1}{2}\epsilon_1\epsilon_2  + \frac{1}{2}\epsilon_2\epsilon_3 \biggr).\nonumber \\
\label{eq:effectivepot}
\end{eqnarray}
Despite the appearance of the slow-roll parameters, this expression is exact. As usual, 
the initial conditions on the perturbations are imposed when the modes
are well inside the Hubble radius during inflation. In this regime, the modes 
do not feel spacetime curvature and, consequently, are
usually chosen to be in the Bunch-Davies vacuum. This amounts to
demanding that the Mukhanov-Sasaki variable $v_{\bm k}$ reduces to
following Minkowski-like positive frequency mode in the sub-Hubble
limit:
\begin{equation}
  \lim_{k/(a\,H)\rightarrow \infty} 
  v_{\bm k}=\f{1}{\sqrt{2\, k}}\; {\rm e}^{-i\,k\,\eta}.
\label{eq:bd-ic}
\end{equation}

\begin{figure}
\includegraphics[width=0.45\textwidth,height=.45\textwidth]{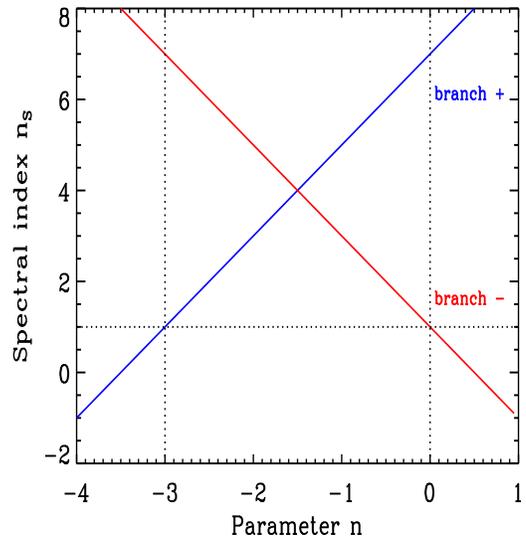}
\caption{Spectral index versus parameter $n$ for the new family 
of potentials.}
\label{fig:spectralns}
\end{figure}

In ultra slow-roll inflation, using Eq.~(\ref{eq:ultrasrparam}), the 
effective potential for the perturbations can be expressed as
\begin{equation}
\frac{z''}{z}\simeq \frac{1}{\eta^2}\left(2+3n+n^2\right),
\end{equation}
where $\eta $ denotes the conformal time. We see that the solution 
to the mode equation can still be expressed as a Bessel function as it 
the case in the conventional situation. The result reads
\begin{equation}
v_{\bm k}(\eta)=-\frac{1}{2} {(-\pi \eta)}^{\frac{1}{2}} 
{\rm e}^{in\pi/2} H_{n+3/2}^{(1)}(-k\eta),
\end{equation}
where $H_{\nu}^{(1)}(z)$ is the Hankel function of first type. Then, 
for $n<-3/2$, the power spectrum on large Hubble scales can be written as
\begin{equation}
\label{eq:usrps}
{\cal P}_{\zeta}(k)=\frac{H^2}{\pi \epsilon_1\Mpl^2}
\left(\frac{k}{aH}\right)^{2n+6}{\cal F}_{_{\rm USR}}(n),
\end{equation}
where 
\begin{equation}
{\cal F}_{_{\rm USR}}(n)\equiv \frac{2^{-2n-7}}{\Gamma^2(n+5/2)\cos^2(n\pi)}.
\end{equation}
To our knowledge, this solution is new although the case $n=-3$ 
was found before in Ref.~\cite{Kinney:2005vj}. If $n>-3/2$, then one has 
\begin{equation}
\label{eq:srps}
{\cal P}_{\zeta}(k)=\frac{H^2}{\pi \epsilon_1\Mpl^2}
\left(\frac{k}{aH}\right)^{-2n}{\cal F}_{_{\rm SR}}(n),
\end{equation}
where 
\begin{equation}
{\cal F}_{_{\rm SR}}(n)\equiv \frac{2^{-1+2n}}{\Gamma^2(-n-1/2)\cos^2(n\pi)}.
\end{equation}
Finally, it remains the case $n=-3/2$. One finds 
\begin{equation}
{\cal P}_{\zeta}(k)=\frac{H^2}{\pi \epsilon_1\Mpl^2} 
\left(\frac{k}{aH} \right)^3 \frac{1}{4\pi^2}\ln^2 \left(\frac{k}{aH}\right),
\end{equation}
In all these expressions (and this is of course crucial for the 
case $n<-3/2$), $\epsilon_1$ must be evaluated not at the time
of Hubble radius  crossing but at the time of consideration, 
typically the end of inflation (of course, in the slow-roll case, 
this does not make a difference since the slow-roll parameters 
remain small and constant). The above expressions lead to the following 
spectral index for the power spectrum
\begin{align*}
n_{_{\rm S}}-1 & = \begin{cases}
                 2(n+3), & n<-3/2 \label{eq:ns}\\
                 -2n, & n>-3/2 \\
                 3+2\ln ^{-1}\left[k/(aH)\right],  & n=-3/2.
           \end{cases}
\end{align*}
The spectral index versus the parameter $n$ is represented 
in Fig.~\ref{fig:spectralns}. One sees that scale invariance is achieved 
for two values, namely $n\simeq 0$ which corresponds to the usual 
slow-roll and $n\simeq -3$ which corresponds to ultra slow-roll. If 
$n\lesssim -3$ the spectrum is red while if 
$n\gtrsim -3$, it is blue. It is easy to check that $0.96<n_{_{\rm S}}<1$,
see Ref.~\cite{Komatsu:2010fb}, 
corresponds to $-3.02<n<-3$. Therefore, we obtain a new family of 
solutions leading to an almost scale invariant power spectrum but, clearly, $n$ 
cannot deviate from $-3$ too strongly. One can also re-express the spectral 
index in terms of the slow-roll parameters. For the slow-roll regime one 
obtains $n_{_{\rm S}}=1-\epsilon_2$ while for the ultra slow-roll regime 
one has
\begin{equation}
n_{_{\rm S}}=2n+6=7+\epsilon_2.
\end{equation}
This should be compared to the standard slow-roll formula, 
$n_{_{\rm S}}=1-2\epsilon_1-\epsilon_2$. Of course, in the slow-roll 
regime, we obtain exactly the same equation given that $\epsilon_1\ll 1$. 
In the ultra slow-roll regime, however, we observe a breakdown of this 
result. This was already noticed in Ref.~\cite{Kinney:2005vj} for the 
case $n=-3$ and it was shown in that reference that this is due to 
a breakdown of the horizon crossing formalism. Indeed, for $n=-3$, the 
slow-roll formalism leads to $n_{_{\rm S}}=7$ instead of the correct 
result $n_{_{\rm S}}=1$.

\par

\begin{figure}
\includegraphics[width=0.45\textwidth,height=.45\textwidth]{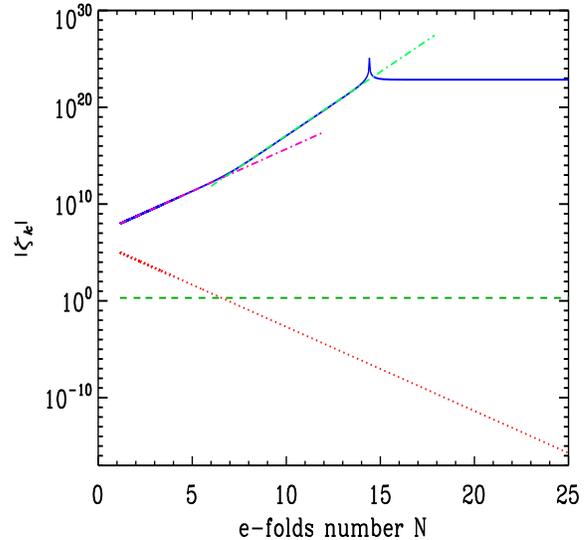}
\caption{Exact (numerical) evolution of curvature perturbations (solid blue line) versus 
the number of e-folds. The dotted dashed pink line represents the scaling 
$\propto a^{-(n+1)}$ while dotted dashed green line correspond to the 
scaling $\propto a^{-(2n+3)}$. The dotted red line is the quantity $k^2/(aH)^2$ 
for a mode such that $k/a_{\rm ini}\sim 50 H_{\rm ini}$ at the beginning of inflation.
The dashed green line represents the quantity $\eta^2 z''/z$.}
\label{fig:modefunction}
\end{figure}

It is also interesting to discuss in more detail the behavior 
of curvature perturbations on large scales. During inflation, the super-Hubble
condition $k/(a\,H)\ll 1$ amounts to neglecting the $k^2$ term with
respect to the effective potential $z''/z$ in the differential
equation~(\ref{eq:eomv}).  In such a case, it is straightforward to
show that the super-Hubble solution to $v_{\bm k}$ can be written as
follows
\begin{eqnarray}
  v_{\bm k}(\eta) 
  &\simeq& A_{\bm k}\, z(\eta)+\, B_{\bm k}\, z(\eta)\, \int ^{\eta}
  \f{\d{\bar \eta}}{z^2({\bar \eta})},
\label{eq:sh-solution}
\end{eqnarray}
where $A_{\bm k}$ and $B_{\bm k}$ are $k$-dependent constants that are
determined by the Bunch-Davies initial condition~(\ref{eq:bd-ic})
chosen in the sub-Hubble limit. In our case, it is easy to show that 
this reduces to 
\begin{equation}
\zeta_{\bm k}\propto A_{\bm k}+B_{\bm k}\, a^{-(2n+3)}.
\end{equation}
In the slow-roll regime, the first term represents the growing 
mode while the second one corresponds to the decaying one. In 
the ultra slow-roll regime however, the second term dominates 
over the first one (A similar situation was also studied in Refs.~\cite{Seto:1999jc,Saito:2008em}). 
This implies in particular that the power 
spectrum is still a time-dependent quantity on super-Hubble scales 
contrary to the standard case where it is conserved. This is apparent 
in Eq.~(\ref{eq:usrps}) where the $\epsilon_1$ term in the denominator 
is a time-dependent quantity. On the contrary, the same factor in 
Eq.~(\ref{eq:srps}) is constant in time and, as a consequence, the 
slow-roll power spectrum does not evolve on large scales. It 
is also worth mentioning that curvature perturbations grow on 
sub-Hubble scales as well. Indeed since 
$\zeta _{\bm k}\sim v_{\bm k}/(a\sqrt{\epsilon_1})$ and since 
$\vert v_{\bm k}\vert $
stays constant in this case, this immediately implies $\vert \zeta_{\bm k}\vert 
\propto a^{-(n+1)}$. In the slow-roll case, curvature perturbations decreases 
$\propto a^{-1}$.

\par

In order to check these considerations, we have numerically integrated Eq.~(\ref{eq:eomv}). The result is presented in Fig.~\ref{fig:modefunction}. The modulus of curvature perturbations corresponds to the solid blue line. The effective potential for the perturbations $\eta^2 z''/z$ is the dashed green line while 
$k^2/(a^2H^2)$ is the dotted red line. When the dotted red line is 
above the dashed green one, the mode is within the Hubble radius and when it is below, the mode is outside the Hubble radius. In Fig.~\ref{fig:modefunction}, we see that the mode starts its evolution deep inside 
the Hubble radius and crosses it out around $N\simeq 6$. We verify 
that, inside the Hubble radius, $\vert \zeta _{\bm k}\vert $ grows like $a^{-(n+1)}$, this particular scaling being represented by the dotted-dashed pink line. When the mode crosses out the Hubble radius, it is apparent that the behavior of $\vert \zeta_{\bm k}\vert$ is modified. The dotted dashed green line represents the scaling $a^{-(2n+3)}$ and one sees in the figure that it is indeed the scaling of $\vert \zeta _{\bm k}\vert $. Therefore, our numerical integration confirms that, in ultra slow-roll inflation, curvature perturbations grow on small and large scales.
Around $N \simeq 13$, ultra slow-roll inflation comes to an end 
and, as a consequence, the growth of $\zeta_\vk$ stops. 
Then, as clearly seen in the figure, $\zeta_\vk$ stays constant as usual in the slow-roll regime on large scales. 

\par

This continuous growth of curvature perturbations during ultra slow-roll inflation turns out to have 
important physical implications. Since ${\cal P}_{\zeta}(k)$ is a 
time-dependent quantity even on large scales, this means that the amplitude of the power spectrum at the time when inflation ends 
must now be compared with the WMAP normalization (in the slow-roll case, it is 
sufficient to normalize the power spectrum when the modes of cosmological 
interest today leaves the Hubble radius during inflation). If $n\simeq -3$, the power spectrum of the curvature perturbation at the
time when inflation ends is given by
\begin{equation}
{\cal P}_\zeta=\frac{1}{24\pi^2 \epsilon_{1*}}e^{6\Delta N_*} {\left( \frac{M}{\Mpl} \right)}^4,
\end{equation}
where $\Delta N_* \simeq 50-60$ is the number of e-fold between the Hubble radius crossing time
of the relevant mode and the end of inflation. The quantity $\epsilon_{1*}$ is $\epsilon_1$ evaluated at the Hubble radius crossing time.
From the WMAP normalization ${\cal P}_\zeta =2.4 \times 10^{-9}$, see 
Ref.~\cite{Komatsu:2010fb}, we find that for $\Delta N_*=60$, $M$ must satisfy
\begin{equation}
\frac{M}{\Mpl}=7 \times 10^{-42} {\left( \frac{\epsilon_{1*}}{0.01} \right)}^{1/4},
\end{equation}
which is far below the Big Bang Nucleosynthesis (BBN) bound $M > {\cal O}({\rm MeV})$. The 
result is expected. The quantity $\vert \zeta _{\bm k}\vert $ grows 
so much during ultra slow-roll inflation that, in order to match the 
correct level of Cosmic Microwave Background (CMB) fluctuations, one must compensate by a tiny 
mass scale in the potential. Let us 
notice that we also implicitly assume that, after inflation, 
the growth of $\vert \zeta_{\bm k}\vert $ stops. In addition, 
the above estimate is very conservative because it is expressed in 
terms of $\epsilon_{1*}$. Since $\epsilon_1$ is decreasing from the 
beginning of inflation, it is likely that $\epsilon_{1*}\ll 1$. In 
other words, instead of $\Delta N_*$, the constraint could also be 
written in terms of the total number of e-folds. This means that a physically relevant ultra slow-roll inflation model can
last only for a much shorter period than the 60 e-folds usually required.

\section{Ultra Slow-Roll Non Gaussianity}
\label{sec:usrNG}

Let us now turn to the calculation of the three-point correlation 
function. For the case $n=-3$, the calculation was done for the first time in Ref.~\cite{Namjoo:2012aa}. Here we generalize this result for an arbitrary value of the parameter $n$. As is well-known, for slow-roll single field inflation 
with a standard kinetic term, the level of non-Gaussianity is very small, 
of the order of the slow-roll parameters, see Refs.~\cite{Gangui:1993tt,Gangui:1994yr,Wang:1999vf,Gangui:1999vg,Gangui:2000gf,Gangui:2002qc}. This 
result is still true for ultra slow-roll inflation but, now, one of the 
slow-roll parameters is of order one. Therefore, one expects a $\fnl$ parameter of order one as well. We will see that this is what happened although, as 
noticed in Ref.~\cite{Namjoo:2012aa}, the relation 
between $\fnl$ and $n_{_{\rm S}}$ is modified.

\par

The scalar bi-spectrum $\cB_{_{\rm S}}(\vka,\vkb,\vkc)$ is defined in
terms of the three point correlation functions of the Fourier modes of
the curvature perturbation $\cR$ as
follows~\cite{Larson:2010gs,Komatsu:2010fb}:
\begin{eqnarray}
\langle {\hat \cR}_{\vka}\, 
{\hat \cR}_{\vkb}\, {\hat \cR}_{\vkc}\rangle 
&=&\l(2\,\pi\r)^3\; \cB_{_{\rm S}}(\vka,\vkb,\vkc)\nn\\
& &\times\,\delta^{(3)}\l(\vka+\vkb+\vkc\r).\label{eq:bs-d}
\end{eqnarray}
For convenience, we shall set
$ G(\vka,\vkb,\vkc)
=(2\,\pi)^{9/2}\, \cB_{_{\rm S}}(\vka,\vkb,\vkc) $.
Using the Maldacena formalism~\cite{Maldacena:2002vr}, 
the quantity $G(\vka,\vkb,\vkc)$
can be expressed as~\cite{Seery:2005wm,Chen:2005fe,Chen:2010xka} (recall 
that the function $f_{\vk}$ below is the mode function that appears in front 
of the annihilation and creation operators in the canonical decomposition 
of the operator $\hat{\zeta}$)
\begin{widetext}
\begin{eqnarray}
G(\vka,\vkb,\vkc)
&\equiv & \sum_{C=1}^{7}\; G_{_{C}}(\vka,\vkb,\vkc)
\nonumber \\
&\equiv & \Mpl^2\; \sum_{C=1}^{6}\; 
\biggl[f_{\ka}(\ef)\, f_{\kb}(\ef)\,f_{\kc}(\ef) 
\cG_{_{C}}(\vka,\vkb,\vkc)
+\,f_{\ka}^{\ast}(\ef)\, f_{\kb}^{\ast}(\ef)\,
f_{\kc}^{\ast}(\ef)
\, \cG_{_{C}}^{\ast}(\vka,\vkb,\vkc)\biggr]
\nonumber \\ & &
+ G_{7}(\vka,\vkb,\vkc),
\end{eqnarray}
where $\ef$ denotes the final time when the bi-spectrum is to be
evaluated.  The quantities $\cG_{_{C}}(\vka,\vkb,\vkc)$ with $C
=1,\cdots, 6$ are described by the
integrals~\cite{Seery:2005wm,Chen:2005fe,Chen:2010xka}
\begin{eqnarray}
\cG_{1}(\vka,\vkb,\vkc)
&=& 2\,i\,\int_{\ei}^{\ef} \d\eta\; a^2\, 
\epsilon_{1}^2\; \bigl(f_{\ka}^{\ast}\,f_{\kb}'^{\ast}\,
f_{\kc}'^{\ast}+\,{\rm two~permutations}\bigr),\label{eq:cG1}\\
\cG_{2}(\vka,\vkb,\vkc)
&=&-\,2\,i\;\l(\vka\cdot \vkb + {\rm two~permutations}\r)
\,\int_{\ei}^{\ef} \d\eta\; a^2\, 
\epsilon_{1}^2\, f_{\ka}^{\ast}\,f_{\kb}^{\ast}\,
f_{\kc}^{\ast},\label{eq:cG2}\\
\cG_{3}(\vka,\vkb,\vkc)
&=&-\,2\,i\,\int_{\ei}^{\ef} \d\eta\; a^2\,
\epsilon_{1}^2
\, \Biggl[\l(\f{\vka\cdot\vkb}{\kb^{2}}\r)\,
f_{\ka}^{\ast}\,f_{\kb}'^{\ast}\, f_{\kc}'^{\ast}
+\, {\rm five~permutations}\Biggr],\label{eq:cG3}\\
\cG_{4}(\vka,\vkb,\vkc)
&=&i\,\int_{\ei}^{\ef} \d\eta\; a^2\,\epsilon_{1}\,
\epsilon_{2}'\,
\,\bigl(f_{\ka}^{\ast}\,f_{\kb}^{\ast}\,
f_{\kc}'^{\ast}+{\rm two~permutations}\bigr),\label{eq:cG4}\\
\cG_{5}(\vka,\vkb,\vkc)
&=&\frac{i}{2}\,\int_{\ei}^{\ef} \d\eta\;
a^2\, \epsilon_{1}^{3}
\,\Biggl[\l(\f{\vka\cdot\vkb}{\kb^{2}}\r)\,
f_{\ka}^{\ast}\,f_{\kb}'^{\ast}\, f_{\kc}'^{\ast}
+\, {\rm five~permutations}\Biggr],\label{eq:cG5}\\
\cG_{6}(\vka,\vkb,\vkc) 
&=&\frac{i}{2}\,\int_{\ei}^{\ef} \d\eta\; a^2\, 
\epsilon_{1}^{3}
\,\Biggl\{\l[\f{\ka^{2}\,\l(\vkb\cdot\vkc\r)}{\kb^{2}\,\kc^{2}}\r]\, 
f_{\ka}^{\ast}\, f_{\kb}'^{\ast}\, f_{\kc}'^{\ast}
+\, {\rm two~permutations}\Biggr\},\label{eq:cG6}
\end{eqnarray}
\end{widetext}
where $\ei$ denotes the time when the modes $f_{\bm k}$ are well
inside the Hubble radius during inflation. The additional, seventh
term $G_{7}(\vka,\vkb,\vkc)$ arises due to a field redefinition, and
its contribution to $G(\vka,\vkb,\vkc)$ is found to be
\begin{eqnarray}
G_{7}(\vka,\vkb,\vkc)
&=&\left[ \frac{\epsilon_{2}}{2}-2(2n+3) \right]\,
\biggl[\vert f_{\kb}(\ef)\vert^{2}\, 
\vert f_{\kc}(\ef)\vert^{2}\nn\\
&+& \, {\rm two~permutations}\biggr].\label{eq:G7}
\end{eqnarray} 
In the ultra slow-roll case, since $\epsilon_1$ is very tiny while 
$\epsilon_2={\cal O}(1)$,
the above equations show that
$G_7$ gives the dominant contribution to the bi-spectrum for any
configuration of the triangle formed by the vectors $\vka$, $\vkb$ and $\vkc$.
Notice that the second term $-2(2n+3)$ in Eq.~(\ref{eq:G7})
is absent in the standard slow-roll case. This originates from the 
fact that the terms in the cubic action that must be removed by 
field redefinition are of the form $a\epsilon_2\zeta ^2/2+2\zeta \zeta'/H+\cdots $, where the dots denote terms that always involve a spatial derivative of the curvature perturbation. In the standard case, only the first 
term is important because of the conservation
of curvature perturbations on super-Hubble scales.
On the other hand, in the present case where the decaying mode dominates 
over the growing mode, the second term also contributes since $\zeta'\neq 0$~\cite{Namjoo:2012aa}.
It is actually this second term that leads to the violation of the
standard non-Gaussianity consistency relation. Then, the bi-spectrum becomes
\begin{eqnarray}
\cB_{_{\rm S}}(\vka,\vkb,\vkc)&=&-\frac{3}{4}\left( n+2 \right)\frac{(2\pi)^{-1/2}}{k_1^3k_2^3k_3^3}\nonumber \\ 
& & \times \left[ k_3^3P_\zeta (k_1) P_\zeta (k_2)+2~{\rm perms} \right].
\end{eqnarray}
Interestingly enough, the bi-spectrum is of the local type not only
in the squeezed limit but also for any other set of $(\vka,\vkb,\vkc)$.
Then, from the above expression, we can immediately read $\fnl$ which is given by\footnote{
We use the same $\fnl$ as the one used by WMAP.
Notice that Refs.~\cite{Maldacena:2002vr,Chen:2005fe,Martin:2011sn,Hazra:2012yn} use a different sign convention.}
\begin{equation}
\fnl^{_{\rm USR}}=-\frac{5}{2}(n+2).
\end{equation}
As noticed in Ref.~\cite{Namjoo:2012aa}, this gives a relation between $\fnl$ and $n$ different from the Maldacena consistency relation which yields $\fnl ^{\rm sq}=5(1-n_{_{\rm S}})/12\simeq 5n/6$. Finally, it is also interesting to provide a relation between $\fnl$ and $n_{_{\rm S}}$:
\begin{equation}
\fnl^{_{\rm USR}}=\frac{5}{4} \left(3-n_{_{\rm S}}\right),
\end{equation}
where we emphasized again the fact that it is valid for any 
configuration, not only in the squeezed limit. This clearly shows 
that $\fnl$ becomes of order one even if the power-spectrum is almost scale invariant. Such a signal would be marginally detectable by the Planck 
satellite which, in principle, can see $\vert \fnl\vert \gtrsim 5$. Finally, 
let also mention that, in order for the bi-spectrum we have just calculated to describe the non-Gaussianity which would actually be observed in the sky, it is necessary to assume that the growth of $\zeta_{\bm k}$ stops after the end of inflation and that reheating will not modify the result. The latter seems very reasonable as recently shown in Ref.~\cite{Hazra:2012kq}.
 
\section{Discussion and Conclusions}
\label{sec:conclusion}

Let us now recap our main results. Ultra slow-roll is not new and was studied 
in Ref.~\cite{Kinney:2005vj}. It is characterized by a situation where the first 
horizon flow parameter is very small but the second one is of order one. In this 
paper, we have generalized the 
ultra slow-roll regime to a one parameter family models. We have seen 
that, in ultra slow-roll inflation, the curvature perturbation can be dominated by the decaying mode. Despite this property, the corresponding power 
spectrum remains scale invariant and, hence, in agreement with 
the CMB observations. This leads to the interesting 
situation where $\fnl$ is of order one even in a single field model 
with a standard kinetic term. This clearly violates the Maldacena consistency 
relation.

\par

However, ultra slow-roll inflation appears to be plagued with many 
difficulties. Firstly, the system is unstable and the ultra slow 
roll solution is left after a few e-folds only unless one artificially 
fine tunes the initial conditions. Secondly, the continuous growth of 
curvature perturbations implies that the mass scale of the potential 
must be extremely small in order to match the observed level of CMB anisotropy. In fact the corresponding value of $M$ turns out to be unphysical. There is also a third difficulty that we now discuss. As is well-known, when the potential is very flat, the 
quantum effects can dominate over the classical dynamics.
In ultra slow-roll inflation, the typical variation of the 
scalar field (during one e-fold) due to the classical 
dynamics can be expressed as
\begin{equation}
\Delta \phi_{\rm cl}\simeq \frac{-3\Mpl^2}{3+n}\frac{V_\phi}{V}.
\end{equation}
On the other hand, typical quantum jumps are given by 
$\Delta \phi_{\rm quant}\simeq H/(2\pi)$. Therefore, the classical 
equations of motion are valid only if $\Delta \phi_{\rm cl}\gg \Delta \phi_{\rm quant}$. 
Using the previous considerations, this leads to 
\begin{equation}
\frac{\phi}{\Mpl}\gg \frac{M^2}{2\pi \vert n\vert  \sqrt{3}\Mpl^2}.
\end{equation}
Given the requirement~(\ref{eq:ftic}), one can have $60$ e-folds 
of ultra slow-roll inflation in the classical regime only if 
\begin{equation}
\frac{M^2}{2\pi \vert n\vert  \sqrt{3}\Mpl^2}<
\frac{1}{\vert n\vert}
\sqrt{\frac{54 \Delta _{\rm cri}}{\vert n+3\vert}}{\rm e}^{-90},
\end{equation}
that is to say
\begin{equation}
\frac{M}{\Mpl}\lesssim \left(\frac{648 \pi^2 \Delta _{\rm cri}}
{\vert n+3\vert}\right)^{1/4}{\rm e}^{-45}.
\end{equation}
For $\Delta _{\rm cri}=0.1$ and $\vert n+3\vert =0.01$, this gives
\begin{equation}
M\lesssim 1.1\, \mbox{GeV}.
\end{equation}
This is larger than the BBN bound $M > {\cal O}({\rm MeV})$ but 
remains rather small. As we have seen the WMAP normalization provides
a much tighter constraint on $M$. Nevertheless, it is likely that in a 
realistic realization of ultra slow-roll inflation the quantum effects 
play a dominant role.

\par

It seems therefore difficult to produce $60$ e-folds of inflation 
in the ultra slow-roll regime. One can wonder whether the very flat 
region of the potential could only represents a limited part of 
the full potential. It seems however difficult to understand how the field 
could enter this part of the potential with the correct initial conditions 
$\ddot{\phi}=nH\dot{\phi}$. Of course if $V_\phi=0$ exactly, then the 
previous condition is true but this does not represents a realistic 
case as there will always be corrections, even if extremely small. 
In this case, moreover, the dynamics would be completely controlled by quantum effects. 

\par

In conclusion, ultra slow-roll inflation represents an interesting 
playground but it remains a challenge to build a physically relevant 
model that would exhibit in this regime. In fact, this shows how robust 
the Maldacena consistency condition is. In order to violate it, we are forced 
to consider situations that appear to be plagued with many physical 
difficulties.

\begin{acknowledgments}
This work was supported by JSPS Research Fellowships for Young Scientists (H.M.) and 
Grant-in-Aid for JSPS Fellows No.~1008477 (T.S.). J.M. would like to 
thank RESCEU (University of Tokyo) for warm hospitality. We would like to 
thank J.~Yokoyama for careful reading of the manuscript and interesting 
comments.
\end{acknowledgments}

\bibliography{biblio}

\end{document}